\documentclass{article}

\newcommand{\ath}{\ensuremath{\mathop{\mathrm{arctanh}}}}

\newcommand{\cA}{{\cal A}}

\newcommand{\be}{\begin{equation}}
\newcommand{\ee}{\end{equation}}
\newcommand{\bea}{\begin{eqnarray}}
\newcommand{\eea}{\end{eqnarray}}

\begin{document}

\title{\begin{flushright}
\small DTP-MSU-0020 \\ \small \tt
gr-qc/0101101
\end{flushright} \vspace{.5cm}
Solitons In Non-Abelian Born-Infeld Theory \footnote{Contribution to the 9th
Marcel Grossmann meeting (MG9), Rome, July 2000}}

\author{{\large D.V. Gal'tsov\footnote{Supported by RFBR grant 00-02-16306.}\ \ and V.V. Dyadichev$^\dagger$ }\\
\it Department of Theoretical Physics, \\
\it Moscow State University, 119899, Moscow,
Russia\\
E-mail: galtsov@grg.phys.msu.su, rkf@mail.ru}

\maketitle

\begin{abstract} Born-Infeld generalization of the
Yang-Mills action suggested by the superstring theory gives
rise to modification of previously known as well as to some new
classical soliton solutions. Earlier it was shown that within
the model with the usual trace over the group generators
classical glueballs exist which form an infinite sequence
similar to the Bartnik-McKinnon family of the
Einstein-Yang-Mills solutions. Here we give the generalization
of this result to the 'realistic' model with the symmetrized
trace and show the existence of excited monopoles (in presence
of triplet Higgs) which can be regarded as a non-linear
superposition of monopoles and sphalerons. \end{abstract}

An effective action for two coincident D-branes is given by the
non-Abelian $SU(2)$ Born-Infeld action (NBI) with some specific
prescription for the trace over the gauge group generators.
Here we use the symmetrized trace following the Tseytlin's
proposal. We find closed expressions for this action for
configurations usually invoked in the search of soliton
solutions: monopole purely magnetic spherically symmetric
ansatz, its dyon generalization, axially symmetric
multimonopole ansatz used, (non-self-dual) instanton ansatz as
well as their self-gravitating versions. In the simplest static
spherically symmetric monopole case with $
A_{\theta}=-(1-w)T_3,\; A_{\varphi}=\sin\theta(1-w)T_3$ the
flat-space action obtained reads \bea \label{cAdef}
L_{NBI}&=&\frac{\beta^2}{4\pi}\left(1-\frac{1+V^2+K^2\cA}
{\sqrt{1+V^2}} \right)\nonumber\\
\cA&=&\sqrt{\frac{1+V^2}{V^2-K^2}}
\ath\sqrt{\frac{V^2-K^2}{1+V^2}}, \eea where \be
V^2=\frac{(1-w^2(r))^2}{2\beta^2 r^4},\quad
K^2=\frac{w'^2(r)}{2\beta^2 r^2}, \ee and $\beta$ is the BI
critical field. The (triplet) Higgs field was introduced under
the standard form. The resulting lagrangian has the following
soliton solutions:
\begin{enumerate}
\item
an infinite sequency of purely Yang-Mills glueballs, half of
which are of sphaleronic nature,
\item
non-BPS magnetic monopoles,
\item
hybrid monopole-glueball solutions.
\end{enumerate}
For all of them the expansion of $w(r)$ in the vicinity of the
origin reads \be\label{wor} w=1-br^2+ O(r^4), \ee where $b$ is
a free parameter. At infinity one has \be\label{was} w=\pm
(1-\frac{c}{r})+O(1/r^2) \ee with some constant $c$ for $1)$
and $w=0$ for $2),\, 3)$. The boundary conditions for the Higgs
are standard. These expansions can be matched for the discrete
values of the parameter $b$ depending on $\beta$. In the purely
glueball case one can rescale the radial variable to set
$\beta=1$ then the first three values of $b$ are given in Tab.1.
\begin{table}\begin{center}
\begin{tabular}{|l|l|l|}
  \hline $n$ & $\quad b$                     & $\quad M$
\\\hline $1$ & $\quad 1.23736\times 10^2$    & $\quad 1.20240$
\\\hline $2$ & $\quad 5.05665\times 10^3$    & $\quad 1.234583$
\\\hline $3$ & $\quad 1.67739\times 10^5$    & $\quad 1.235979$
\\ \hline
\end{tabular}
\end{center}
\caption{Discrete values of $b$ and the corresponding masses
$M$ of NBI glueballs with symmetrized trace}
 \end{table}
The integer $n$ gives the number of zeroes of $w$. These values
are about an order of magnitude greater than those obtained in
the NBI theory with an ordinary trace~\cite{GaKe99}.
Qualitatively solutions remain the same, so the symmetrized
action modification relates to the deep core region of
glueballs \cite{DyGa001}. Recall, that these solutions are
unstable for all $n$, those with odd $n$ can be interpreted as
sphalerons. Their mass gives the heights of the potential
barriers separating neighboring topological vacua. With growing
$n$ the masses rapidly converge to the mass of an embedded
abelian solution, which can be regarded as the limiting
solution for large $n$ ( in the weak sense). When gravity is
included~\cite{DyGa002}, these solutions can be related to the
Bartnik-McKinnon solutions of the Einstein-Yang-Mills equations
in the strong gravity limit.

In the case of monopoles one has three dimensional parameters
of which one can be set to unity by rescaling. Choosing as two
remaining parameters $\beta$  and $\lambda$ (the Higgs
potential coefficient), one finds monopole solutions for
$\beta>\beta_{cr}$ with the critical value not very different
from that obtained by Schaposnik et al. In addition we have
found hybrid monopole-glueball solutions which have the
following $w$-behavior. Starting with (\ref{wor}), the function
$w$ crosses zero and becomes negative like in for the $n=1$
sphaleron, but then, instead of approaching the vacuum value
$w=-1$ at infinity, it turns back to monopole regime and
finally $w(\infty)=0$. Similarly, higher $n$ solutions exist
which approach the same asymptotic value after any number of
oscillation around zero ( crossing zero $n$ times). These
solutions have masses higher that the monopole mass and they
are unstable. Again, this is similar to the situation in the
usual YM-Higgs theory coupled to gravity.

\eject
\end{document}